\documentclass[aps,pra,preprint,amsmath,amssymb,floatfix,superscriptaddress,showpacs]{revtex4} 

\usepackage[pdftex]{graphicx} 
\usepackage{bm} 
\usepackage{ulem} 
\usepackage{pifont}

\begin{document}

\title{Effects of Spatially Nonuniform Gain on Lasing Modes in Weakly 
Scattering Random Systems}
\author{Jonathan Andreasen}
\altaffiliation[Also at]{
  Department of Physics and Astronomy, Northwestern University,
  Evanston, IL 60208, USA
}
\affiliation{
  Department of Applied Physics, Yale University,
  New Haven, CT 06520, USA
}
\author{Christian Vanneste}
\affiliation{
  Laboratoire de Physique de la Mati\`ere Condens\'ee, CNRS UMR 6622,
  Universit\'e de Nice-Sophia Antipolis, 
  Parc Valrose, 06108, Nice Cedex 02, France 
}
\author{Li Ge}
\affiliation{
  Department of Applied Physics, Yale University,
  New Haven, CT 06520, USA
}
\author{Hui Cao}
\affiliation{
  Department of Applied Physics, Yale University,
  New Haven, CT 06520, USA
}
\affiliation{
  Department of Physics, Yale University,
  New Haven, CT 06520, USA
}
\date{\today}
\begin{abstract}
  A study on the effects of optical gain nonuniformly distributed in 
  one-dimensional random systems is presented.
  It is demonstrated numerically that even without gain saturation and mode
  competition, the spatial nonuniformity of gain can cause dramatic and
  complicated changes to lasing modes.
  Lasing modes are decomposed in terms of the quasi modes of the passive
  system to monitor the changes.
  As the gain distribution changes gradually from uniform to nonuniform,
  the amount of mode mixing increases.
  Furthermore, we investigate new lasing modes created by nonuniform gain 
  distributions. 
  We find that new lasing modes may disappear together with existing lasing 
  modes, thereby causing fluctuations in the local density of lasing states.
\end{abstract}
\maketitle 

\section{Introduction}

Lasing modes in random media behave quite differently depending on the 
scattering characteristics of the media \cite{caoLRM}.
In the strongly scattering regime, lasing modes have a nearly one-to-one 
correspondence with the localized modes of the passive system 
\cite{VannestePRL01,souk02}.
Due to small mode volume, different localized modes may be selected for lasing 
through local pumping of the random medium \cite{caoprl00,VannestePRL01}.
The nature of lasing modes in weakly scattering open random systems 
(e.g., \cite{frolov,ling01,Mujumdar}) is still under discussion 
\cite{wiersman}.
In systems which are diffusive on average, prelocalized modes may serve as 
lasing modes \cite{apalkov}.
In general, however, the quasi modes of weakly scattering systems are very 
leaky. 
Hence, they exhibit a large amount of spatial and spectral overlap.
For inhomogeneous dielectric systems with uniform gain distributions, even 
linear contributions from gain induced polarization bring about a coupling 
between quasi modes of the passive system \cite{deych05PRL}.
Thus, lasing modes may be modified versions of the corresponding quasi modes.
However, Vanneste \textit{et al.} found that when considering uniformly
distributed gain, the first lasing mode appears to correspond to a single 
quasi mode \cite{VannestePRL}.
The study was done near the threshold pumping rate and nonlinear effects did 
not modify the modes significantly.
Far above threshold, it was found that lasing modes consist of a collection of 
constant flux states \cite{tureciSci}.
Mode mixing in this regime is largely determined by nonlinear effects from 
gain saturation.

Remaining near threshold, pumping a local spatial region, and including
absorption outside the pumped region found lasing modes to differ 
significantly from the quasi modes of the passive system \cite{yamilovOL}.
This change is attributed to a reduction of the effective system size.
More surprisingly, recent experiments \cite{polson,wu06} and numerical studies 
\cite{wu07} showed the spatial characteristics of lasing modes change 
significantly by local pumping even without absorption in the unpumped region.
It is unclear how the lasing modes are changed in this case by local pumping.
In this paper, we carry out a detailed study of random lasing modes in a weakly 
scattering system with a nonuniform spatial distribution of linear gain.
Mode competition depends strongly on the gain material properties, 
e.g., homogeneous vs. inhomogeneous broadening of the gain spectrum.
Ignoring gain saturation (usually responsible for mode competition) and 
absorption, we find that spatial nonuniformity of linear gain alone can 
cause mode mixing.
We decompose lasing modes in terms of quasi modes and find them to be a 
superposition of quasi modes close in frequency.
The more the gain distribution deviates from being uniform, more quasi modes 
contribute to a lasing mode.

Furthermore, still considering linear gain and no absorption outside the gain 
region, we find that some modes stop lasing no matter how high the gain is.
We investigate how the lasing modes disappear and further investigate the 
properties of new lasing modes \cite{andreasenOL} that appear.
The new lasing modes typically exist for specific distributions of gain and 
disappear as the distribution is further altered. 
They appear at various frequencies for several different gain distributions.

In Section \ref{sec:method}, we describe the numerical methods used to study 
the lasing modes of a one-dimensional (1D) random dielectric structure.
The model of gain and a scheme for decomposing the lasing modes in terms of 
quasi modes is presented. 
A method to separate traveling wave and standing wave components from the total 
electric field is introduced.
The results of our simulations are presented and discussed in 
Section \ref{sec:results}. 
Our conclusions concerning the effects of nonuniform gain on lasing modes 
are given in Section \ref{sec:conclusion}.

\section{Numerical Method\label{sec:method}}

The 1D random system considered here is composed of $N$ layers.
Dielectric layers with index of refraction $n>1$ alternate with air gaps 
($n=1$) resulting in a spatially modulated index of refraction $n(x)$.
The system is randomized by specifying different thicknesses for each of the
layers as
$d_{1,2} = \left<d_{1,2}\right>(1+\eta\zeta)$ where
$\left<d_1\right>$ and $\left<d_2\right>$ are the average
thicknesses of the layers, $0 < \eta < 1$ represents the 
degree of randomness, and $\zeta$ is a random number in (-1,1).

A numerical method based on the transfer matrix is used to calculate both the 
quasi modes and the lasing modes in a random structure.
Electric fields on the left (right) side of the structure $p_0$ ($q_N$) and
$q_0$ ($p_N$) travel toward and away from the structure, respectively.
Propagation through the structure is calculated via the $2\times 2$ matrix $M$
\begin{equation}
  \left(
  \begin{array}{c}
    p_N \\
    q_N \\
  \end{array}
  \right)=
  M
  \left(
  \begin{array}{c}
    p_0 \\
    q_0 \\
  \end{array}
  \right). \label{eq:transfermatrix}
\end{equation}
The boundary conditions for outgoing fields only are $p_0=q_N=0$, requiring 
$M_{22}=0$.

For structures without gain, wavevectors must be complex in order to satisfy 
the boundary conditions.
The field inside the structure is represented by 
$p(x)\exp[i n(x) \tilde{k} x] + q(x)\exp[-i n(x) \tilde{k} x]$, 
where $\tilde{k}=k+ik_i$ is the complex 
wavevector and $x$ is the spatial coordinate.
For outgoing-only boundary conditions ($M_{22}=0$), $k_i < 0$.
The resulting field distributions associated with the solutions for these 
boundary conditions are the quasi modes of the passive system.

Linear gain is simulated by appending an imaginary part to the dielectric
function $\epsilon(x)=\epsilon_r(x)+i\epsilon_i(x)$, where 
$\epsilon_r(x)=n^2(x)$.
This approximation is only valid at or below threshold \cite{souk99}.
In Appendix \ref{ap:lineargain}, the complex index of refraction is calculated 
as $\tilde{n}(x) = \sqrt{\epsilon(x)} = n_r(x) + in_i$, where $n_i < 0$.
We consider $n_i$ to be constant everywhere within the random system. 
This yields a gain length $l_g=1/k_i=1/n_ik$ ($k=2\pi/\lambda$ is the vacuum
wavenumber of a lasing mode) which is the same in the dielectric layers and 
the air gaps.
Note that this gain length is not to be confused with the length of the spatial 
gain region, which is described below.
The real part of the index of refraction is modified by the imaginary part
as $n_r(x)=\sqrt{n^2(x)+n_i^2}$.
A real wavevector $k = 2\pi/\lambda$ describes the lasing frequency.
The field inside the structure is now represented by
$p(x)\exp[i \tilde{n}(x) k x] + q(x)\exp[-i \tilde{n}(x) k x]$.
The frequencies and thresholds are located by determining which values 
of $k$ and $n_i$, respectively, satisfy $M_{22}=0$.

Nonuniform gain is introduced through an envelope function $f_E(x)$
multiplying $n_i$.
The envelope considered here is the step function $f_E(x)=H(-x+l_G)$, 
where $x=0$ is the left edge of the random structure and $x=l_G$ is the 
location of the right edge of the gain region.
$l_G$ may be chosen as any value between 0 and $L$.

The solutions of the system are given by the points at which the complex 
transfer matrix element $M_{22}=0$.
Where $\operatorname{Re}[M_{22}]=0$ or $\operatorname{Im}[M_{22}]=0$, 
``zero lines'' are formed in the plane of either complex $\tilde{k}$ 
(passive case) or ($k$, $n_i$) (active case).
The crossing of a real and imaginary zero line results in $M_{22}=0$ at that 
location, thus revealing a solution.
We visualize these zero lines by plotting 
$\log_{10}\left|\operatorname{Re}M_{22}\right|$ and 
$\log_{10}\left|\operatorname{Im}M_{22}\right|$ to enhance the regions near 
$M_{22}=0$ and using various image processing techniques to enhance the contrast.

The solutions are pinpointed more precisely by using the Secant method.
Locations of minima of $|M_{22}|^2$ and a random value located closely to these 
minima locations are used as the first two inputs to the Secant method.
Once a solution converges 
or $|M_{22}| < 10^{-12}$, a solution is considered found.
This method has proved extremely adept at finding genuine solutions when 
suitable initial guesses are provided.

Verification of these solutions is provided by the phase of $M_{22}$, 
calculated as 
$\theta=\operatorname{atan2}(\operatorname{Im}M_{22},\operatorname{Re}M_{22})$.
Locations of vanishing $M_{22}$ give rise to phase singularities since both 
the real and imaginary parts of $M_{22}$ vanish. 
The phase change around a path surrounding a singularity in units of $2\pi$ 
is referred to as topological charge \cite{halperin,ZhangJOSAA}. 
In the cases studied here, the charge is $+ 1$ for phase increasing in 
the clockwise direction and $- 1$ for phase increasing in the 
counterclockwise direction.

Once a solution is found, the complex spatial field distribution may be 
calculated. 
Quasi modes $\psi(x)$ are calculated in the passive case and lasing modes 
$\Psi(x)$ are calculated in the active case.
In order to determine whether or not mode mixing occurs (and if so, to what 
degree) in the case of nonuniform gain, the lasing modes are decomposed
in terms of the quasi modes of the passive system.
It was found \cite{LeungJPA,LeungJPA2} that any spatial function 
defined inside an open system of length $L$ [we consider the lasing modes 
$\Psi(x)$ here] can be expressed as
\begin{equation}
  \Psi(x) = \sum_m a_m \psi_m(x),\label{eq:recon}
\end{equation}
where $\psi_m(x)$ are a set of quasi modes characterized by the complex 
wavevectors $\tilde{k}_m$.
The coefficients $a_m$ are calculated by
\begin{align}
  a_m =& i \int_0^L\left[\Psi(x)\hat{\psi}_m(x)+\hat{\Psi}(x)\psi_m(x)\right]dx 
  \nonumber\\
  &+ i\left[\Psi(0)\psi_m(0)+ \Psi(L)\psi_m(L)\right],\label{eq:coeffs}
\end{align}
where
\begin{subequations}
  \begin{equation}
    \hat{\psi}_m(x)=-i\tilde{k}_m n^2(x)\psi_m(x)\label{eq:conjmoma}
  \end{equation}
  \begin{equation}
    \hat{\Psi}(x)=-ik\left[n_r(x)+in_if_E(x)\right]^2\Psi(x).\label{eq:conjmomb}
  \end{equation}
\end{subequations}
The normalization condition is
\begin{align}
  1 =&i \int_0^L 2\psi(x)\hat{\psi}(x) dx 
  + i\left[\psi^2(0)+ \psi^2(L)\right]\nonumber\\
  =& i \int_0^L 2\Psi(x)\hat{\Psi}(x) dx
  + i\left[\Psi^2(0)+ \Psi^2(L)\right].\label{eq:normalization}
\end{align}

An advantage of this decomposition method is that a calculation over an 
infinite system has been reduced to a calculation over a finite system.
Error checking is done by using the coefficients found in 
Eq. (\ref{eq:coeffs}) to reconstruct the lasing mode intensity distribution 
with Eq. (\ref{eq:recon}) yielding $R(x) \equiv \sum_m a_m \psi_m(x)$.
We define a reconstruction error $E_R$ to monitor the accuracy of the 
decomposition:
\begin{equation}
  E_R = \frac{\int | \Psi(x) - R(x)|^2 dx}{\int |\Psi(x)|^2 dx}.
\end{equation}

In general, the field distribution of a mode in a leaky system consists of 
a traveling wave component and a standing wave component.
In Appendix \ref{ap:sttr} we introduce a method to separate the two.
This method applies to quasi modes as well as lasing modes; we shall consider
lasing modes here.
At every spatial location $x$, the right-going complex field $\Psi^{(R)}(x)$ 
and left-going complex field $\Psi^{(L)}(x)$ are compared. 
The field with the smaller amplitude is used for the standing wave and the 
remainder for the traveling wave.
If $|\Psi^{(R)}(x)| < |\Psi^{(L)}(x)|$ then the standing wave component 
$\Psi^{(S)}(x)$ and traveling wave component $\Psi^{(T)}(x)$ are
\begin{subequations}
  \begin{equation}
    \Psi^{(S)}(x) = \Psi^{(R)}(x) + [\Psi^{(R)}(x)]^*
  \end{equation}
  \begin{equation}
    \Psi^{(T)}(x) = \Psi^{(L)}(x) - [\Psi^{(R)}(x)]^*.
  \end{equation}
  \label{eq:standtrav}
\end{subequations}

Further physical insight on lasing mode formation and disappearance, as well as
new lasing mode appearance is provided by a mapping of an 
``effective potential'' dictated by the random structure.
Local regions of the random medium reflect light at certain frequencies but 
are transparent to others \cite{kuhl08}.
The response of a structure to a field with frequency $\omega=ck$ can be
calculated via a wavelet transformation of the real part of the dielectric 
function $\epsilon_r(x) = n^2(x)$ \cite{bliojosab}.
The relationship between the local spatial frequency $q_{res}$ and the optical
wavevector $k$ is approximately $q_{res}=2k$ in weakly scattering structures.
The Morlet wavelet $\chi$ is expressed as
\begin{equation}
  \chi\left(\frac{x'-x}{s}\right) = 
  \frac{\pi^{-1/4}}{\sqrt{s}}e^{i\omega_0(x'-x)/s}
  e^{-(x'-x)^2/2s^2},
\end{equation}
with nondimensional frequency $\omega_0$ and Gaussian envelope
width $s$ \cite{torrencecompo}.
With $\omega_0$ fixed, stretching the wavelet through $s$ changes the 
effective frequency. 
Wavelets with varying widths are translated along the spatial axis 
to obtain the transformation
\begin{equation}
  W(x,q_{res}) = \int \epsilon_r(x') 
  \chi^*\left(\frac{x'-x}{s}\right)dx'.\label{eq:wavelett}
\end{equation}
where
\begin{equation}
  q_{res} = \frac{\omega_0 + \sqrt{2+\omega_0^2}}{2s}.
\end{equation}
The wavelet power spectrum $|W(x,q_{res})|^2$ is interpreted as an effective 
potential.
Regions of high power indicate potential barriers and regions of low
power indicate potential wells for light frequency 
$\omega = q_{res}c/2$.

\section{Results\label{sec:results}}

A random system of $N=161$ layers is examined in the following as an example
of a random 1D weakly scattering system.
The indices of refraction of the dielectric layers are 
$n_1 = 1.05$ and the air gaps $n_2 = 1$.
The average thicknesses are $\left<d_1\right> = 100$ nm and 
$\left<d_2\right> = 200$ nm  giving a total average length of
$\left<L\right> =$ 24100 nm.
The grid origin is set at $x=0$ and the length of the random structure $L$ is 
normalized to $\left<L\right>$.
The degree of randomness is set to $\eta = 0.9$ and the index of
refraction outside the random media is $n_0 = 1$.
The localization length $\xi$ is calculated from the dependence of 
transmission $T$ of an ensemble of random systems of different
lengths as $\xi^{-1} = -d\left<\ln T\right>/dL$.
The above parameters ensure that the localization length is nearly constant 
at 200 $\mu$m $\le \xi \le 240$ $\mu$m over the wavelength range 
500 nm $\le \lambda \le$ 750 nm.
With $\xi \gg L$, the system is in the ballistic regime.

\begin{figure}
  \centering
  \includegraphics[width=8.5cm]{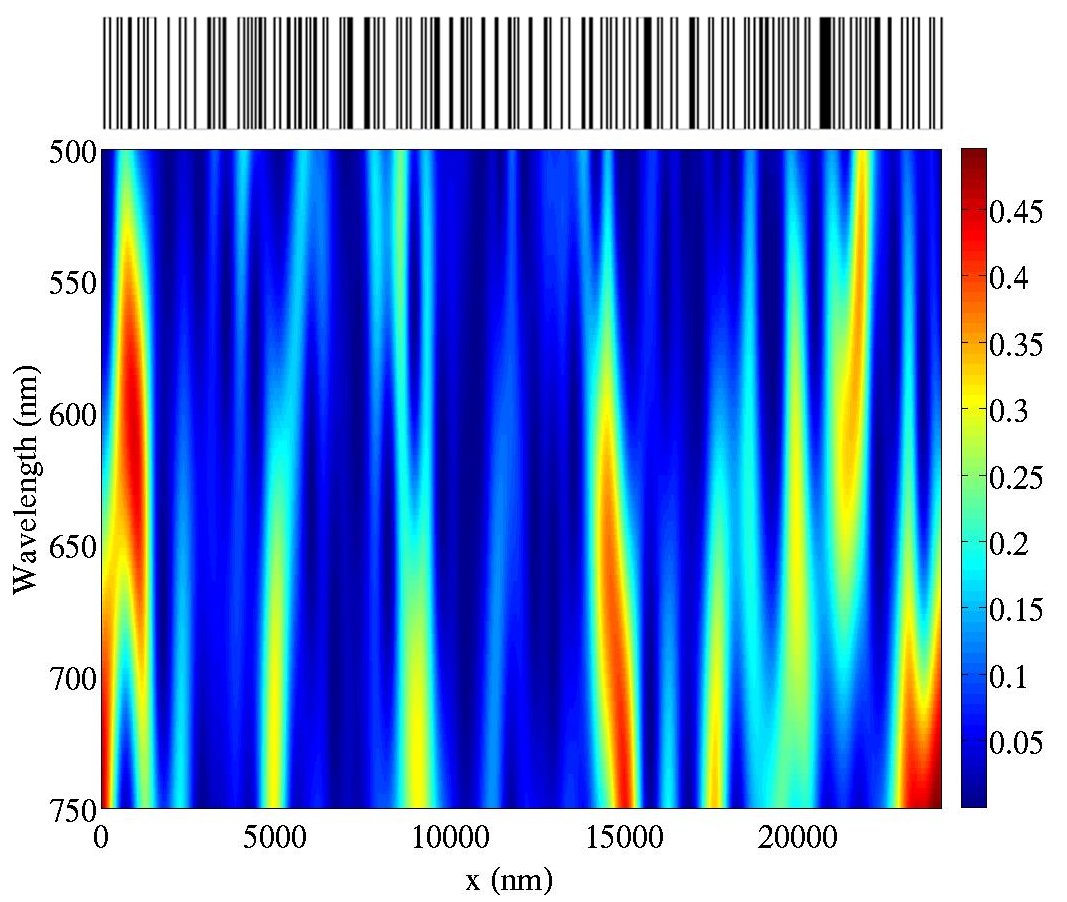}
  \caption{\label{fig:wavelet} (Color online)
    Effective potential (wavelet power spectrum) $|W(x)|^2$ of the dielectric 
    function $\epsilon_r(x)$ as a function of position $x$ and 
    wavelength $\lambda$.
    Regions of high power indicate potential barriers and regions of low
    power indicate potential wells where intensities are typically
    trapped.
    The black lines on the top represent the spatial distribution of dielectric
    constant $\epsilon_r(x)=n^2(x)$.
  }
\end{figure}

Figure \ref{fig:wavelet} shows the effective potential of the structure within 
the wavelength range of interest via a wavelet transformation.
We use a nondimensional frequency of $\omega_0=6$ \cite{farge} and a spatial 
sampling step of $\Delta x = 2$ nm.
The power spectrum $|W(x)|^2$ reveals the landscape of the effective potential
dictated by the locations and thicknesses of the dielectric layers.

\begin{figure}
  \includegraphics[width=8.5cm]{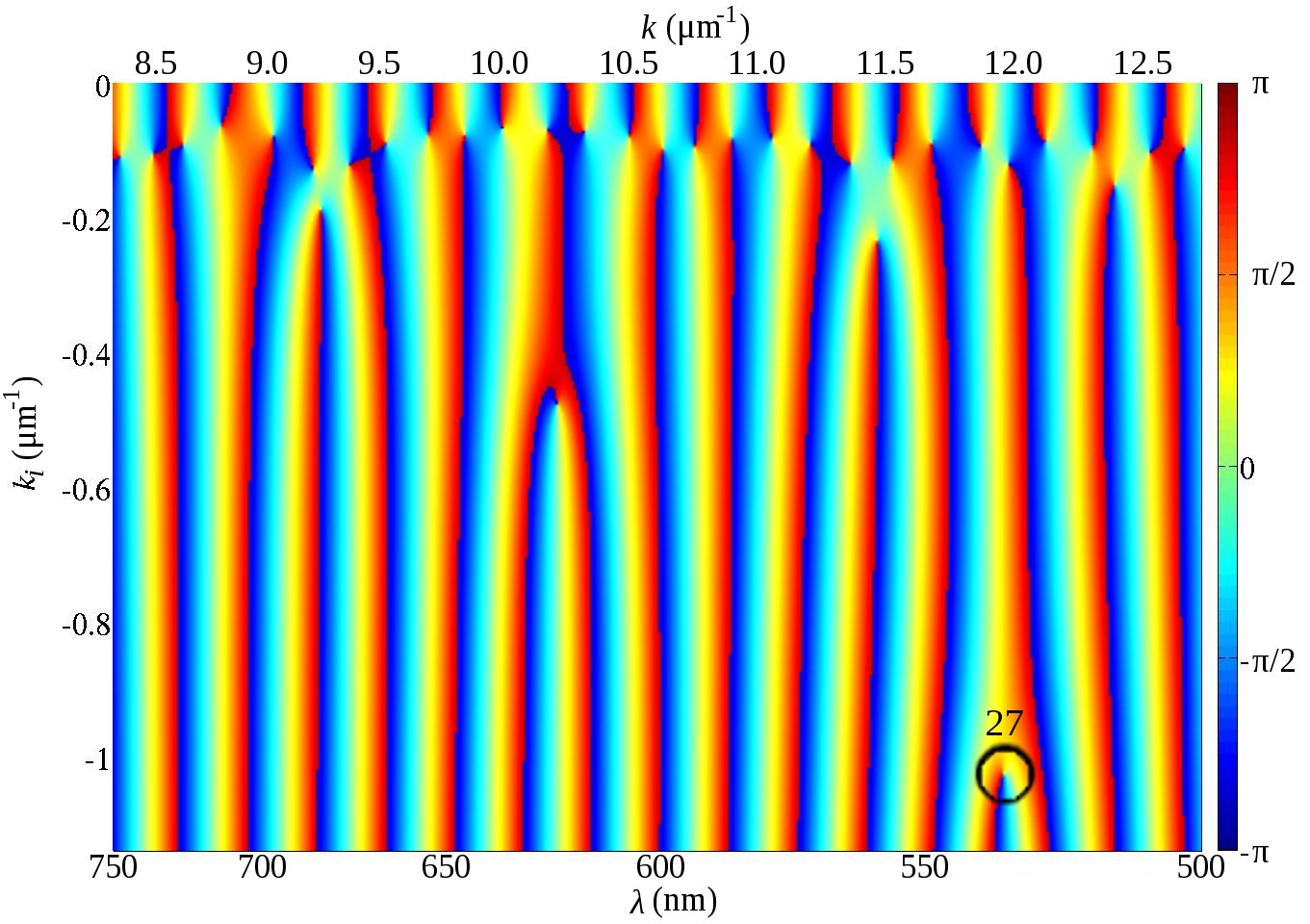}
  \caption{\label{fig:phasemap161} (Color online)
    A mapping of the phase $\theta$ of $M_{22}$ for the passive 1D random
    system without gain.
    The topological charge of all quasi modes seen here is $-1$. 
    Modes are enumerated from left to right.
    Quasi mode 27 is encircled in black.
  }
\end{figure}

Figure \ref{fig:phasemap161} is a phase map of $M_{22}$ in the passive case
(without gain).
The phase singularities mark the quasi modes' $\tilde{k}$ values and are 
indicated by phase changes from $-\pi$ to $\pi$ along any lines passing through.
The topological charge of all quasi modes is $-1$. 
Adjacent modes are formed by real and imaginary zero lines of $M_{22}$ that are
not connected to one another. 
We calculated $M_{22}$ for increasingly large $|k_i|$ values until machine 
precision was reached and no additional modes appeared.
As previously found \cite{wu07}, mode frequency spacing is fairly constant in 
the ballistic regime.
The nearly equal spacing of phase singularities in Fig. \ref{fig:phasemap161} 
attests to this.

\begin{figure}
  \includegraphics[width=8.5cm]{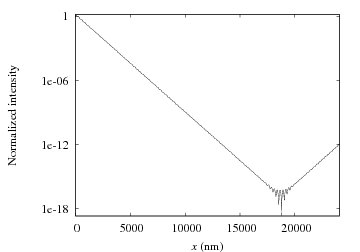}
  \caption{\label{fig:wf27} 
    Normalized intensity $|\psi(x)|^2$ of a leaky mode--quasi mode 27 
    ($\tilde{k} = 11.8 - i 1.03$ $\mu$m$^{-1}$) 
    of the passive random 1D structure.
    The intensity is peaked at the left boundary of the structure,
    similar to doorway states \cite{oko03}.
  }
\end{figure}

Most quasi modes have similar decay rates except for a few which have 
much larger decay rates.
Modes are enumerated here starting with the lowest frequency mode in our
wavelength range of interest. 
Mode 1 has a wavelength of 748 nm and mode 33 has a wavelength of 502 nm.
Most quasi modes have $k_i$ values around $-0.1$ $\mu$m$^{-1}$.
But a few have much larger decay rates, such as mode 27 at $\lambda=532$ nm
which has $k_i=-1.03$ $\mu$m$^{-1}$ (encircled in black in 
Fig. \ref{fig:phasemap161}).
Figure \ref{fig:wf27} shows the intensity of mode 27 to be concentrated 
on one side of the open structure.
We observe that it bears similarity to ``doorway states'' common to 
open quantum systems \cite{oko03}.
Doorway states are concentrated around the boundary of a system and strongly 
couple to the continuum of states outside the structure. 
Therefore, they have much larger decay rates.

For the case of uniform gain, only the lasing modes with large thresholds 
change significantly from the quasi modes of the passive system.
Finding the corresponding quasi modes for lasing modes with large thresholds is
challenging due to changes caused by the addition of a large amounts of gain.
Thus, we neglect them in the following comparisons.
However, there is a clear one-to-one correspondence with quasi modes for the 
remaining lasing modes. 
The average percent difference between quasi mode frequencies and lasing mode
frequencies is 0.026\%.
The average percent difference between quasi mode decay rates $k_i$ and lasing 
thresholds (multiplied by $k$ for comparison) $n_ik$ is 4.3\%.
The normalized intensities of the quasi modes $I_Q(x)\equiv |\psi(x)|^2$ and 
lasing modes $I_L(x)\equiv |\Psi(x)|^2$ are also compared.
The spatially averaged percent difference between each pair of modes is
calculated as
$(2/L)\int\left\{ |I_{Q}(x)-I_{L}(x)|/[I_{Q}(x)+I_{L}(x)]\right\}dx\times 100$.
The averaged difference between intensities of the 3 lasing modes with the
largest thresholds (of the large threshold modes not neglected) compared to 
the quasi modes is 68\% while the remaining pairs average a difference of 4.0\%.

\subsection{Nonuniform Gain Effects on Lasing Mode Frequency, Threshold, 
and Intensity Distribution}

\begin{figure}
  \includegraphics[width=8.5cm]{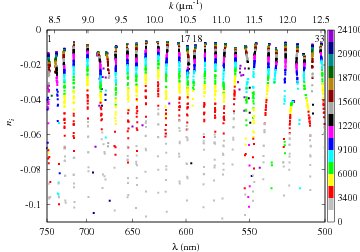}
  \caption{\label{fig:follow} (Color online) 
    Frequencies and thresholds ($k$, $n_i$) of the lasing modes of the 1D 
    random structure with gain. 
    Lasing modes 1, 17, 18, and 33 are explicitly marked.
    The gain region length $l_G$ reduces from uniform gain ($l_G=L$) to 
    nonuniform gain $l_G < L$.
    The color indicates the value of $l_G$ (units of nm) decremented along the 
    layer interfaces.
    Due to the random thicknesses of the layers, the $l_G$ decrements are 
    unequal.
    Hence, the reason for the unequal spacing of the color code.
  }
\end{figure}

Figure \ref{fig:follow} maps the ($k$, $n_i$) values of lasing modes as 
nonuniform gain is introduced by reducing the gain region length from $l_G=L$.
In this weakly scattering system the intensity distributions of modes are 
spatially overlapping. 
This results in a repulsion of mode frequencies \cite{kramer93}. 
As the size of the gain region changes, the envelopes of the intensity 
distributions change, but for most modes $n_i$ is small enough to leave the 
optical index landscape unchanged. 
Thus, the modes continue to spatially overlap as the size of the gain region 
changes and their frequencies remain roughly the same as in the uniform gain 
case. 
Similar behavior of lasing mode frequencies can be seen as the gain region 
length is varied in a simpler cavity with uniform index.
Thus, the robustness of frequency is not due to inhomogeneity in the spatial
dielectric function.
However, the threshold values of the lasing modes change as $l_G$ decreases.
Due to the limited spatial region of amplification, the thresholds increase.
The increase of $n_i$ due to the change of threshold, though considerable, is 
not large enough to significantly impact the lasing frequencies as evidenced by
the small change of frequencies as $l_G$ decreases.

\begin{figure}
  \includegraphics[width=8.5cm]{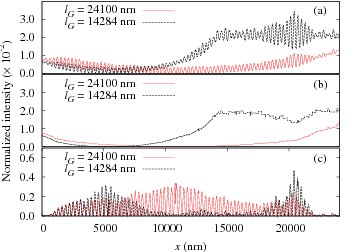}
  \caption{\label{fig:wfs1} (Color online)
    Normalized intensity of lasing mode 17 with uniform gain $l_G=24100$ nm
    (red solid lines) and nonuniform gain $l_G=14284$ nm (black dashed lines). 
    Gain is only located in the region $0 \le x \le l_G$.
    (a) Total intensity $|\Psi(x)|^2$, (b) traveling wave intensity 
    $|\Psi^{(T)}(x)|^2$, and (c) standing wave intensity 
    $|\Psi^{(S)}(x)|^2$.
    Nonuniform gain significantly changes the spatial intensity envelope as 
    well as the standing wave and traveling wave components.
  }
\end{figure}

The intensity distributions of the lasing modes also change considerably as 
$l_G$ is reduced.
Normalized spatial intensity distributions are given by $|\Psi(x)|^2$ 
after $\Psi(x)$ has been normalized according to Eq. (\ref{eq:normalization}).
The intensities are sampled with a spatial step of $\Delta x = 2$ nm.
With uniform gain ($l_G=24100$ nm), the intensity of lasing mode 17 
($\lambda = 598$ nm) in Fig. \ref{fig:wfs1}(a) increases toward the gain 
boundaries due to weak scattering and strong amplification.
When the gain boundary is changed to $l_G=14284$ nm, the envelope of the spatial
intensity distribution changes dramatically.
The intensity increases more rapidly toward the boundaries of the gain region
and stays nearly constant outside the gain region but still inside the 
structure.
This change can be understood as $|n_i|$ inside the gain region causes the 
intensity to become larger, while outside the gain region $n_i=0$ and the 
wavevector is real.

To monitor the change in the trapped component of the intensity, $\Psi(x)$ is
separated into a traveling wave and a standing wave component via 
Eq. (\ref{eq:standtrav}) (see Appendix \ref{ap:sttr}).
Figures \ref{fig:wfs1}(b) and (c) show the traveling wave and standing wave 
components of lasing mode 17, respectively.
For $l_G=L$, the intensity increase toward the structure boundaries is caused
mostly by the growth of the traveling wave.
The standing wave part is strongest near the center of the system. 
For $l_G=14284$ nm, the standing wave exhibits two peaks, one concentrated 
near the center of the gain region and the other outside the gain region.
However, the standing wave intensity outside the gain region should not be 
directly compared to the standing wave intensity inside the gain region. 
The total intensity inside the gain region increases toward the gain boundary 
in this weakly scattering system.
Thus, the amplitude of the field outside the gain region, where there is no 
amplification, is determined by the total field amplitude at the gain boundary.
The randomness of the dielectric function outside the gain region traps part
of the wave which results in a relatively large standing wave intensity 
compared to inside the gain region.
However, outside the gain region, there is a net flux toward the right 
boundary of the system meaning the traveling wave intensity in this region is 
large as well. 

\begin{figure}
  \includegraphics[width=8.5cm]{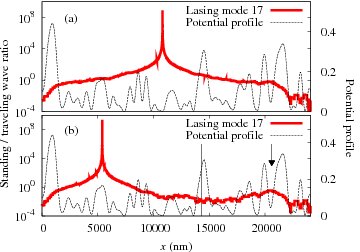}
  \caption{\label{fig:wfs2} (Color online)
    Standing/traveling wave ratio $A_{ST}(x)$ (red solid lines) of lasing 
    mode 17 with uniform gain $l_G=L$ (a) and nonuniform gain 
    $l_G= 14284$ nm (b).
    Gain is located in the regions left of the vertical black solid line.
    We term the location at which the ratio diverges 
    ($A_{ST}(x)\rightarrow\infty$), the standing wave center (SWC).
    The potential profile $|W(x)|^2$ (black dashed lines) of the dielectric 
    function at the wavelength of mode 17 is overlaid in both (a) and (b).
  }
\end{figure}

The relative strength of the standing wave is calculated through the ratio of 
standing wave amplitude to traveling wave amplitude. 
The amplitudes are calculated in Appendix \ref{ap:sttr}.
Depending on whether the prevailing wave is right-going or left-going, the
standing/traveling wave ratio is given by
\begin{equation} 
  A_{ST}(x) = 
  \left\{
    \begin{array}{l}
      \left|\frac{2\Theta(x)}{\Pi(x)-\Theta(x)}\right|^2, \Pi(x) > \Theta(x)\\
      \left|\frac{2\Pi(x)}{\Theta(x)-\Pi(x)}\right|^2, \Pi(x) < \Theta(x).
    \end{array}
    \right.
\end{equation}
Results from considering uniform and nonuniform gain for lasing mode 17 are 
shown in Fig. \ref{fig:wfs2}.
Where the standing wave is largest inside the gain region, 
$|\Psi^{(T)}(x)|=|\Theta(x)-\Pi(x)|=0$ 
and the ratio $A_{ST}(x)$ is infinite.
The location where the ratio is diverging is the position of pure standing wave.
Fields are emitted in both directions from this position.
The prevailing wave to the right of this standing wave center (SWC) is 
right-going.
The prevailing wave to the left of this SWC is left-going.
The SWC of the lasing mode is located near the center of the total system when 
considering uniform gain in Fig. \ref{fig:wfs2}(a).
With the size of the gain region reduced in Fig. \ref{fig:wfs2}(b), we see 
that the SWC of the lasing mode (where $A_{ST}(x)\rightarrow \infty$) moves to 
stay within the gain region. 
Furthermore, note that this mode has a relatively small threshold (see 
Fig. \ref{fig:follow}).
We have found that in general, modes with low thresholds have a SWC near the 
center of the gain region while high threshold modes have a SWC near the edge 
of the gain region.

The cause for the small peak of $A_{ST}(x)$ outside the gain region can be 
found in the potential profile of Fig. \ref{fig:wavelet}.
A slice of the potential profile $|W(x)|^2$ at the wavelength of mode 17 
($\lambda = 598$ nm) is overlaid on the intensities in Fig. \ref{fig:wfs2}.
This suggests the standing wave is weakly trapped in a potential well around 
$x=20500$ nm [marked by an arrow in Fig. \ref{fig:wfs2}(b)].

\subsection{Mode Mixing}

Lasing modes can be expressed as a superposition of quasi modes of the passive
system via Eq. (\ref{eq:recon}) for any distribution of gain.
Coefficients obtained from the decomposition of the lasing modes in terms of 
the quasi modes by Eq. (\ref{eq:coeffs}) offer a clear and quantitative way to 
monitor changes of lasing modes by nonuniform gain.
Using Simpson's rule for the numerical integrations and a basis consisting of 
at least 15 quasi modes at both higher and lower frequencies than the lasing 
mode being decomposed, we consistently find $E_R \approx 10^{-4}$.

\begin{figure}
  \includegraphics[width=8.5cm]{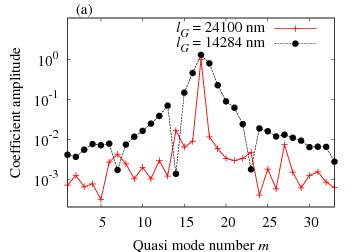}
  \includegraphics[width=8.5cm]{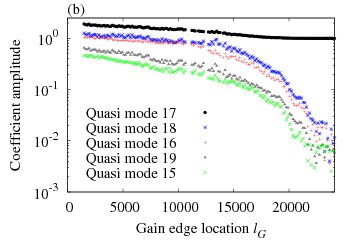}
  \caption{\label{fig:c1732} (Color online)
    Decomposition of lasing mode 17 in terms of passive quasi modes.
    (a) Decomposition with uniform gain (red crosses) and nonuniform 
    gain (black circles).
    Leaky quasi modes, i.e., modes with large $|k_i|$ such as modes
    7, 14, and 23, contribute to lasing modes markedly different than the 
    others.
    (b) Five largest coefficients from the decomposition of lasing mode 17.
    As $l_G$ is reduced, the amount of mode mixing increases dramatically.
    The reconstruction error $E_R$ for lasing mode 17 is close to $10^{-4}$ 
    until $l_G = 11000$ nm then rises to $10^{-2}$ at $l_G=3200$ nm.
    Some coefficients are greater than one, which is possible in open systems.
  }
\end{figure}

Figure \ref{fig:c1732}(a) shows the decomposition of lasing mode 17 
with uniform and nonuniform gain.
Beginning with the case of uniform gain ($l_G=L$), the largest 
contribution to lasing mode 17 is from corresponding quasi mode 17.
There is a nonzero contribution from other quasi modes on the order $10^{-3}$. 
This reflects slight differences between the lasing mode profile in the 
presence of uniform gain and the quasi mode profile \cite{deych05PRL,wu07}.
With the gain region length reduced to $l_G=14284$ nm, the coefficients 
$|a_m|^2$ from quasi modes closer in frequency to the lasing modes increase 
significantly; i.e., quasi modes closer in frequency are mixed in. 
The exceptions are the very leaky quasi modes 7, 14, and 23.
Unlike leaky quasi mode 27 shown in Fig. \ref{fig:wf27}, quasi modes 7, 14, and
23 have intensities which are peaked at the right boundary of the structure.
It has been observed that when $l_G$ reduces and the intensity distribution of 
lasing mode 17 shifts to the left boundary of the structure, there is less 
overlap with these leaky quasi modes.
Thus, the magnitude of the coefficients associated with the leaky modes reduces
as shown in Fig. \ref{fig:c1732}(a).

Figure \ref{fig:c1732}(b) reveals the five largest coefficients $|a_m|^2$ for 
lasing mode 17 as $l_G$ is incrementally reduced along the dielectric 
interfaces.
While the lasing mode remains dominantly composed of its corresponding quasi 
mode, neighboring quasi modes mix into the lasing mode significantly.
It was shown \cite{deych05PRL} that linear contributions from gain induced 
polarization bring about a coupling between quasi modes of the passive system.
This coupling arises solely due to the inhomogeneity of the dielectric 
function, not the openness of the system.
While this interaction may play a role in mode mixing with uniform 
gain, the effect is small compared to the mode mixing caused by the 
nonuniformity of the gain.
This is clearly demonstrated in Fig. \ref{fig:c1732}(b), where the 
coefficients from quasi modes close in frequency are orders of magnitude 
larger for small $l_G$ than for $l_G=L$.

\subsection{Lasing Mode Disappearance and Appearance}

\begin{figure*} 
  \includegraphics[width=6.0cm]{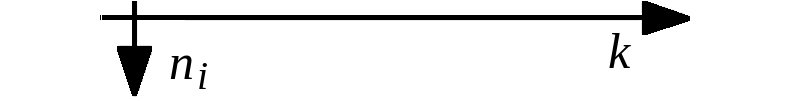}
  \hspace{2cm}
  \includegraphics[width=6.0cm]{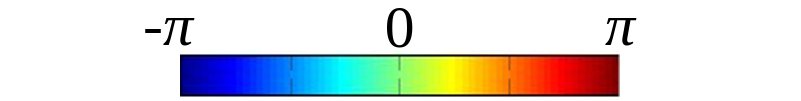}\\
  \includegraphics[width=4.0cm]{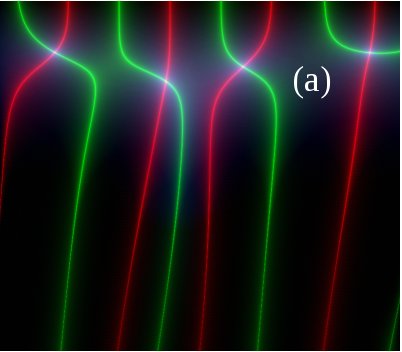}
  \includegraphics[width=4.0cm]{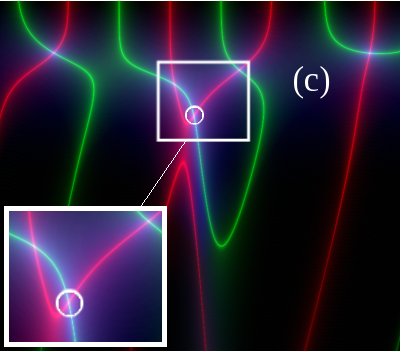}
  \includegraphics[width=4.0cm]{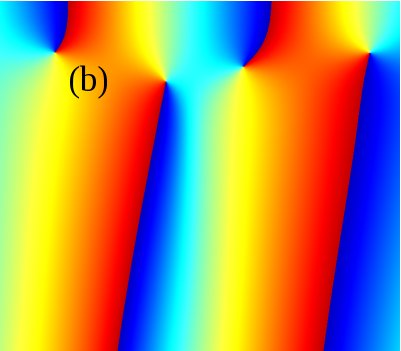}
  \includegraphics[width=4.0cm]{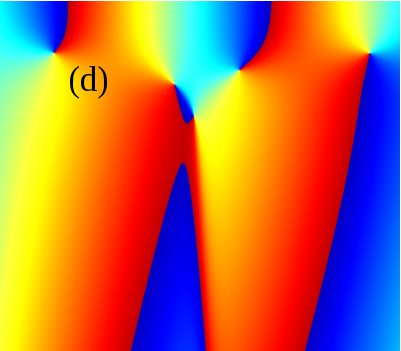}\\
  \includegraphics[width=4.0cm]{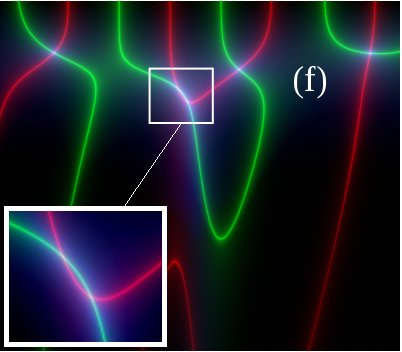}
  \includegraphics[width=4.0cm]{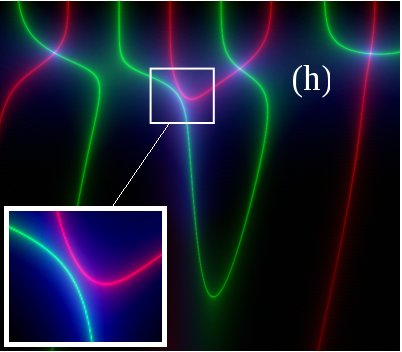}
  \includegraphics[width=4.0cm]{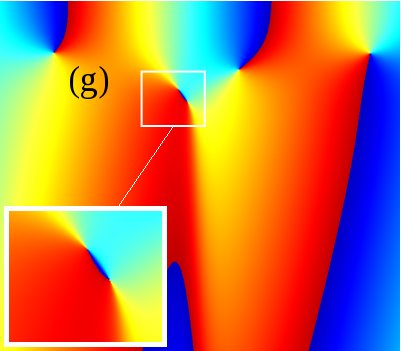}
  \includegraphics[width=4.0cm]{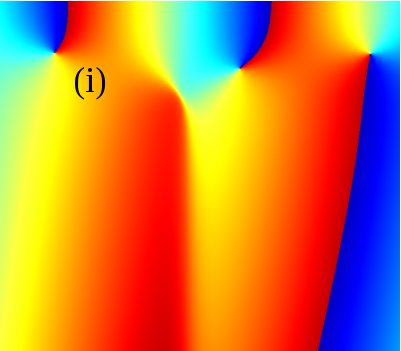}\\
  \includegraphics[width=4.0cm]{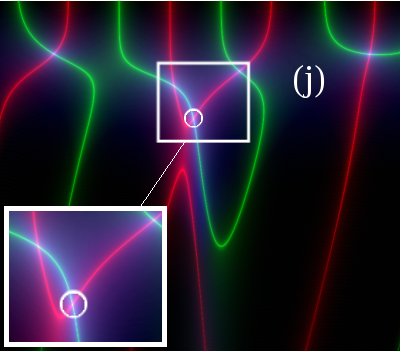}
  \includegraphics[width=4.0cm]{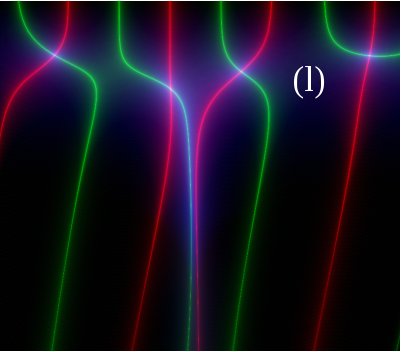}
  \includegraphics[width=4.0cm]{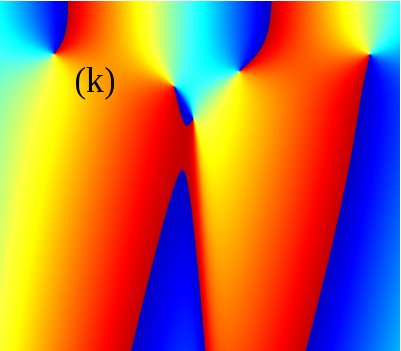}
  \includegraphics[width=4.0cm]{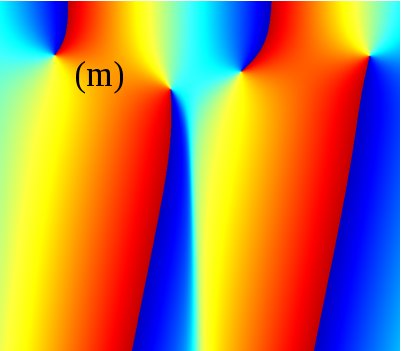}
  \caption{\label{fig:ssezls} (Color online)
    (left) Real (green) and Imaginary (red) zero lines of $M_{22}$. 
    Their crossings indicate ($k$, $n_i$) values of lasing modes.
    (right) Phase maps of $M_{22}$.
    All data are in the ranges
    ($10.3$ $\mu$m$^{-1} < k < 10.8$ $\mu$m$^{-1}$) and 
    ($0 \ge n_i \ge -0.074$)
    covering lasing modes 17 and 18 for $l_G= 14961$ nm (a-b), 14553 nm (c-d), 
    14523 nm (f-g), 14472 nm (h-i), 14284 nm (j-k), and 14042 nm (l-m).
    The joining of zero lines in (c) results in the formation of a new lasing 
    mode (new zero line crossing is encircled in white).
    The inset in (c) is an enlargement of the mode 17 and new mode solutions.
    In (d), the phase singularity at the new mode has a topological charge of 
    $+1$, opposite to that of mode 17.
    The real and imaginary zero lines pull apart in (f) so that the mode 17 and 
    new mode solutions are nearly identical.
    The phase map in (g) reveals the existence of the two phase singularities.
    The lines completely separate in (h) resulting in the disappearance of 
    mode 17 and the new mode.
    The phase map in (i) confirms the disappearance of the two modes.
    This process reverses itself in (j-m) yielding behavioral symmetry around 
    $l_G=14472$ nm.
  }
\end{figure*}

As the size of the gain region reduces we observe that some lasing modes 
disappear and new lasing modes appear.
The existence of new lasing modes in the presence of gain saturation has 
been confirmed \cite{andreasenOL}.
This phenomenon is not limited to random media, but its occurrence has been
observed in a simple 1D cavity with a uniform index of refraction.
New lasing modes, to the best of our knowledge, are always created with larger 
thresholds than the existing lasing modes adjacent in frequency.
The disappearance of lasing modes is not caused by mode competition for gain 
because gain saturation is not included in our model of \textit{linear gain}.
Disappearance/appearance events occur more frequently for smaller values of 
$l_G$.
New lasing modes appear at frequencies in between the lasing mode frequencies 
of the system with uniform gain.
These new modes exist only within small ranges of $l_G$.
We also find that the disappearance events exhibit behavioral symmetry (as
explained below) around particular values of $l_G$. 
This disappearance and subsequent reappearance causes a fluctuation of the 
local density of lasing states as $l_G$ changes.

We examine the progression of one representative event in detail.
The gaps in the decomposition coefficients for lasing mode 17 in 
Fig. \ref{fig:c1732}(b), in the range 10500 nm $\le l_G \le 14500$ nm, indicate
lasing mode 17 does not exist for those distributions of gain.
Figure \ref{fig:ssezls} shows the real and imaginary zero lines of $M_{22}$ 
and their accompanying phase maps for $l_G=$ 14961 nm, 14553 nm, 14523 nm, 
14472 nm, 14284 nm, and 14042 nm.
As $l_G$ decreases, the zero lines of lasing modes 17 and 18 join as seen in 
the transition from Fig. \ref{fig:ssezls}(a) to (c).
This creates a new mode solution (marked by a white circle) with a frequency 
between lasing modes 17 and 18 and a larger threshold.
The existence of a new lasing mode is confirmed by the phase singularity in
Fig. \ref{fig:ssezls}(d).
The new mode is close to mode 17 in the ($k$, $n_i$) plane and its phase 
singularity has the opposite topological charge as seen in 
Fig. \ref{fig:ssezls}(d).
As $l_G$ decreases further, the joined zero lines forming mode 17 and the new 
mode pull apart.
This causes the two solutions to approach each other in the ($k$, $n_i$) 
plane, i.e., the frequency and threshold of mode 17 increase while the
frequency and threshold of the new mode decrease.
In Figs. \ref{fig:ssezls}(f) and (g), the solutions are so close that they are 
nearly identical, yet they still represent two separate solutions.
Further decreasing $l_G$ makes the solutions identical.
The zero lines then separate and the phase singularities of opposite charge 
annihilate each other in Figs. \ref{fig:ssezls}(h) and (i).
This results in the disappearance of mode 17 and the new mode.
The process then reverses itself as $l_G$ is decreased further 
[Figs. \ref{fig:ssezls}(j)-(m)] yielding the reappearance of mode 17 and the new
mode and their subsequent separation in the ($k$, $n_i$) plane. 
This is the aforementioned behavioral symmetry around $l_G=14472$ nm.

\begin{figure}
  \includegraphics[width=8.5cm]{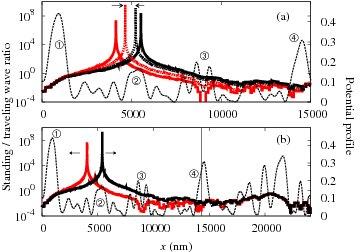}
  \caption{\label{fig:wfs094} (Color online)
    (a) Standing/traveling wave ratio $A_{ST}(x)$ of the new lasing mode 
    (red) and lasing mode 17 (black) for $l_G=14553$ nm (solid lines) and 
    $l_G=14523$ nm (dotted lines).
    The potential profile $|W(x)|^2$ (black dashed line) of the dielectric 
    function for this wavelength is overlaid in both (a) and (b) and major 
    potential barriers are marked \ding{192} through \ding{195}.
    The ratio $A_{ST}(x)$ of the new lasing mode and lasing mode 17 
    become more similar an converge on each other as $l_G$ reduces.
    Reducing $l_G$ further causes these two lasing modes to first disappear
    then reappear as the process reverses itself.
    (b) Standing/traveling wave ratio $A_{ST}(x)$ of the new lasing mode (red)
    and lasing mode 17 (black) after they have reappeared for $l_G=14284$ nm.
    The ratios are similar to the ratios for $l_G=14553$ nm in (a).
    The vertical black solid line marks the gain edge.
    The ratios of the modes diverge now as $l_G$ is reduced.
  }
\end{figure}

Examining the standing/traveling wave ratio of lasing mode 17 and the new lasing
mode together with the potential profile $|W(x)|^2$ offers some insight of 
mode annihilation and reappearance in real space.
Figure \ref{fig:wfs094} shows the ratio $A_{ST}(x)$ for the new mode and
mode 17 along with $|W(x)|^2$.
The potential profile is very similar for the new mode and mode 17 since their 
wavelengths are very close.
There are four major potential barriers at the mode 17 wavelength 
($\lambda = 598$ nm) for $x < 15000$ nm.
This is the spatial region associated with the gain distributions in 
Fig. \ref{fig:ssezls} where $l_G$ is always smaller than 15000 nm.
Figure \ref{fig:wfs094} shows them at: $x = $ \ding{192} 927 nm, 
\ding{193} 5200 nm, \ding{194} 8700 nm, and \ding{195} 14519 nm.
Due to oscillations, the centers of barriers \ding{193} and \ding{194} are 
less well defined.
The right edge of the gain region at $l_G=14553$ nm is located just to the 
right of barrier \ding{195}.
For $l_G=14523$ nm, the right edge of the gain region nears the maximum of 
barrier \ding{195}.
Figure \ref{fig:wfs094}(a) shows that for $l_G=14553$ nm, the SWC of the new 
mode is between barrier \ding{192} and barrier \ding{193}.
The SWC of mode 17 is in the middle of the gain region at $x = 5300$ nm and 
its SWC is between barrier \ding{193} and barrier \ding{194}.
Before disappearing, the modes approach each other in the ($k$, $n_i$) plane, 
eventually merge, and their intensity distributions become identical (as 
evidenced by the trend of their standing/traveling wave ratios).
As $l_G$ is further reduced and the modes reappear, the behavior of the modes' 
ratios $A_{ST}(x)$ (or equivalently, intensity distributions) reverses itself as 
expected from the behavioral symmetry shown in Fig. \ref{fig:ssezls}.
At $l_G=14284$ nm, the right edge of the gain region has passed barrier 
\ding{195} and Fig. \ref{fig:wfs094}(b) [with a different horizontal scale than 
Fig. \ref{fig:wfs094}(a)] shows the SWC of the new mode is in roughly the same 
location as it was for $l_G=14553$ nm.
The SWC of mode 17 is also in roughly the same location as it was for 
$l_G=14553$ nm.

\begin{figure}
  \includegraphics[width=8.5cm]{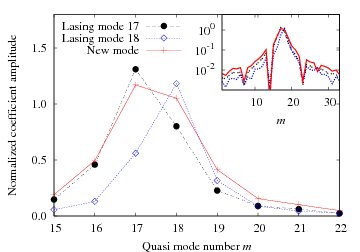}
  \caption{\label{fig:c17nm1nm2} (Color online)
    Decomposition at $l_G=14284$ nm of lasing mode 17 (black circles), lasing 
    mode 18 (blue open diamonds), and the new lasing mode (red crosses) in 
    terms of the quasi modes of the passive system.
    Lasing modes 17 and 18 are mostly composed of their respective quasi modes 
    while the new mode is dominated by a mixture of both quasi mode 17 and 18.
    The inset shows the decomposition coefficients of outlying quasi modes
    for lasing mode 17 (black line), lasing mode 18 (blue line), and the new
    lasing mode (red line).
  }
\end{figure}

The appearance of new lasing modes is unanticipated.
In the passive system, the number of standing wave peaks 
for quasi modes increases incrementally by 1, 
e.g., quasi mode 17 has 82 peaks and quasi mode 18 has 83 peaks.
Lasing modes 17 and 18 behave the same way.
How exactly does a new lasing mode fit into this scheme?
Though closer in frequency and threshold to lasing mode 17, counting the 
total number of standing wave peaks of the new lasing mode yields the 
same number as for lasing mode 18.
However, the new lasing mode is somewhat compressed in the gain region having 
one more peak than lasing mode 18. 
It is decompressed in the region without gain having one less peak than lasing
mode 18.

Comparing the decompositions of the lasing modes in terms of quasi modes helps
reveal the character of the new lasing mode.
Figure \ref{fig:c17nm1nm2} shows the decomposition of the new lasing mode 
together with the decomposition of lasing modes 17 and 18 at $l_G=14284$ nm.
The new mode has a slightly larger coefficient amplitude associated with quasi 
mode 17 than quasi mode 18, but the two amplitudes are nearly equal.
We found that as mode 17 and the new mode solutions approach each other by
varying $l_G$, their coefficient distributions also approach each other until 
becoming equal as expected from Figs. \ref{fig:ssezls} and \ref{fig:wfs094}.

\section{Conclusion\label{sec:conclusion}}

We have demonstrated the characteristics of lasing modes to be strongly 
influenced by nonuniformity in the spatial gain distribution in 1D random
structures.
While the entire structure plays the dominant role in determining the frequency
of the lasing modes, the gain distribution mostly determines the lasing 
thresholds and spatial distributions of intensity.
The gain distribution also appears to be solely responsible for the creation of
new lasing modes.
We have verified the existence of new lasing modes in numerous random 
structures as well as dielectric slabs of uniform refractive index.
A more thorough investigation of the latter will be described in a future 
work.
All of these changes caused by nonuniform gain take place without the influence
of nonlinear interaction between the field and gain medium.
Our conclusion is that nonuniformity of the gain distribution alone is 
responsible for the complicated behavior observed here.

By decomposing the lasing modes in terms of a set of quasi modes of the 
passive system, we illustrated how the lasing modes change.
The contribution of a quasi mode to a lasing mode was seen to depend mostly on 
its proximity in frequency $k$ and the spatial distribution of gain.
The more the gain changed from uniformity, the greater the mixing in of
neighboring quasi modes.
Thus, great care must be taken even close to the lasing threshold when using 
the properties of quasi modes to predict characteristics of lasing 
modes in weakly scattering systems with nonuniform gain or local pumping.

The change of intensity distributions of lasing modes as the size of the gain 
region is varied appears to be general. 
With reduction of the size of the gain region, the peak of the 
standing/traveling wave ratio $A_{ST}(x)$, or the standing wave center (SWC) of
the mode, moves to stay within the gain region. 
Modes with low thresholds have a SWC near the middle of the gain region while 
high threshold modes have a SWC near the edge of the gain region.
Changing the gain distribution thus changes the intensity distributions of 
lasing modes.
The exact modal distributions, however, appear correlated with the potential 
profile. 
In the cases studied here, the new lasing mode and lasing mode 17 lay in
between two large potential barriers.
Decreasing the size of the gain region brought the intensity distributions 
closer together until they disappeared.
These changes took place by varying the edge of the gain region only hundreds
of nanometers.
Thus, even a slight change in the gain distribution may have drastic 
consequences for lasing modes.

\begin{acknowledgments}
  The authors thank Patrick Sebbah, Alexey Yamilov, A. Douglas Stone, and
  Dimitry Savin for stimulating discussions.
  This work was supported partly by the National Science Foundation under Grant 
  Nos. DMR-0814025 and DMR-0808937.
\end{acknowledgments}

\appendix 

\section{Linear Gain Model \label{ap:lineargain}}
In this appendix, we describe the model used to simulate linear gain in a 1D
system.
The gain is linear in the sense that it does not depend on the electromagnetic 
field intensity.
The lasing solutions $\Psi(x)$ must satisfy the time-independent wave equation
\begin{equation}
  \left[\frac{d^2}{dx^2} + \epsilon(x,\omega)k^2\right]\Psi(x) = 0,
\end{equation}
with a complex frequency-dependent dielectric function 
\begin{equation}
  \epsilon(x,\omega) = \epsilon_r(x) + \chi_g(x,\omega),
\end{equation}
where $\epsilon_r(x)=n^2(x)$ is the dielectric function of the non-resonant
background material.
The frequency dependence of $\epsilon_r(x)$ is negligible.
$\chi_g(x,\omega)$, corresponding to the susceptibility of the resonant material,
is given by
\begin{equation}
  \chi_g(x,\omega)=\frac{A_eN_A(x)}{\omega_a^2-\omega^2
    -i\omega\Delta\omega_a},
  \label{eq:chia1}
\end{equation}
where $A_e$ is a material-dependent constant, $N_A(x)$ is the spatially dependent
density of atoms, $\omega_a$ is the atomic transition frequency, and
$\Delta\omega_a$ is the spectral linewidth of the atomic resonance.
Equation (\ref{eq:chia1}) may be simplified by assuming the frequencies of
interest $\omega$ are within a few linewidths of the atomic frequency $\omega_a$,
i.e., 
$\omega^2-\omega_a^2=(\omega+\omega_a)(\omega-\omega_a)
\approx 2\omega_a(\omega-\omega_a)$.
Equation (\ref{eq:chia1}) then reduces to
\begin{equation}
  \chi_g(x,\omega)\approx \frac{iA_eN_A(x)}{\omega_a\Delta\omega_a
    [1+2i(\omega-\omega_a)/\Delta\omega_a]}.
  \label{eq:chia2}
\end{equation}
The frequency-dependent index of refraction is
\begin{align}
  \tilde{n}(x,\omega) =& 
  \sqrt{\epsilon(x,\omega)} = \sqrt{\epsilon_r(x)+\chi_g(x,\omega)} \nonumber\\
  =& n_r(x,\omega) + in_i(x,\omega),
\end{align}
which may then be implemented in the transfer matrix method.
At this point, let us note that only 2 steps are needed to convert this classical
electron oscillator model to real atomic transitions \cite{siegbook}.
First, the radiative decay rate $\gamma_{\parallel}$ may be substituted in to
Eq. (\ref{eq:chia2}) in place of a few constants.
Second, and more importantly, real quantum transitions induce a response
proportional to the population difference density $\Delta N_A$.
Thus, $N_A(x)$ should be replaced by $\Delta N_A$, the difference in population
between the lower and upper energy levels.

Linear gain independent of $\omega$ is obtained by working in the limit
$\omega-\omega_a \ll \Delta\omega_a$, yielding
\begin{equation}
  \chi_g(x)\approx i\frac{A_e\Delta N_A(x)}{\omega_a\Delta\omega_a}
  \label{eq:chia3},
\end{equation}
a purely imaginary susceptibility.
We can make the definition $\chi_g(x)\equiv i\epsilon_i(x)$, where
$\epsilon_i(x)$ is the imaginary part of $\epsilon(x)$.
Note that $\epsilon(x)$ may include absorption [$\epsilon_i>0$] or gain
[$\epsilon_i<0$].
We shall only consider gain here.
The complex frequency-\textit{independent} dielectric function now yields a
frequency-independent index of refraction $\tilde{n}(x)=n_r(x)+in_i(x)$ which may
be expressed explicitly as
\begin{align}
  n_r(x) =& \frac{n(x)}{\sqrt{2}}
  \left[\sqrt{1+\frac{\epsilon_i^2(x)}{n^4(x)}}+1\right]^{1/2}\nonumber\\
  n_i(x) =& \frac{-n(x)}{\sqrt{2}}
  \left[\sqrt{1+\frac{\epsilon_i^2(x)}{n^4(x)}}-1\right]^{1/2}\label{eq:lgni}.
\end{align}
Furthermore, in the main text, we assume $n_i$ to be spatially independent.
Thus, by solving for $n_r(x)$ in terms of $n(x)$ and $n_i$, the
index of refraction used throughout this paper is given by
\begin{align}
  \tilde{n}(x) =& n_r(x) + in_i\nonumber\\
  =& \sqrt{n^2(x)+n_i^2} + in_i.
\end{align}

\section{Standing wave and traveling wave components of the total field
\label{ap:sttr}}
In this appendix, we describe the method that enables one to define a standing 
wave component and a traveling wave component of the field at each point $x$ 
of a 1D system.

For an open structure without gain, the field reads
\begin{equation}
  \psi(x) = p(x)\exp[in(x)\tilde{k}x]+q(x)\exp[-in(x)\tilde{k}x],
\end{equation}
where $\tilde{k}$ is the complex wavevector and $n(x)$ is the index of 
refraction, the value of which alternates between $n(x)=n_1>1$ in dielectric
layers and $n(x)=n_2=1$ in air gaps.
For structures with gain, the field reads
\begin{equation}
  \Psi(x) = p(x)\exp[i\tilde{n}(x)kx]+q(x)\exp[-i\tilde{n}(x)kx],
\end{equation}
where $\tilde{n}(x)=n(x)+in_i$ is the complex index of refraction.
We rewrite both equations in the single form
\begin{equation}
  E(x) = p(x)\exp[i\tilde{K}(x)x]+q(x)\exp[-i\tilde{K}(x)x],\label{eq:EKx}
\end{equation}
where $\tilde{K}(x)=K_r(x)+iK_i(x)$ and $E(x)$ may be either $\psi(x)$ or 
$\Psi(x)$.

For now, we will consider the field within a single layer in order to simplify
the notation.
The following results will be valid within any layer.
Since within a layer, the coefficients $p(x)$, $q(x)$ and the wavevector 
$\tilde{K}(x)$ do not depend on $x$, we rewrite Eq. (\ref{eq:EKx}) as
\begin{equation}
  E(x) = p\exp[i\tilde{K}x]+q\exp[-i\tilde{K}x].\label{eq:EK}
\end{equation}
The complex amplitudes $p$ and $q$ of the right-going and left-going fields,
respectively, can be written as $p=P\exp[i\varphi]$ and $q=Q\exp[i\phi]$ where
$P$ and $Q$ are the real amplitudes which can be chosen positive.
The field becomes
\begin{align}
  E(x) =& P\exp[-K_ix]\exp[i(K_rx+\varphi)]\nonumber\\
  &+  Q\exp[K_ix]\exp[-i(K_rx-\phi)]\nonumber\\
  =& \Pi(x)\exp[i(K_rx+\varphi)] \nonumber\\
  &+\Theta(x)\exp[-i(K_rx-\phi)],\label{eq:tot}
\end{align}
where $\Pi(x)\equiv P\exp[-K_ix]$ and $\Theta(x)\equiv Q\exp[K_ix]$.
Introducing the global phase $\Phi\equiv [\varphi+\phi]/2$ and the difference
$\Delta\equiv [\varphi-\phi]/2$, the field reads 
\begin{align}
  E(x)=&\exp[i\Phi]\{\Pi(x)\exp[i(K_rx+\Delta)]\nonumber\\
  &+ \Theta(x)\exp[-i(K_rx+\Delta)]\} .
\end{align}
Within a single layer, we can set $\Phi=0$ so that the field becomes
\begin{align}
  E(x)=&\Pi(x)\exp[i(K_rx+\Delta)]+\Theta(x)\exp[-i(K_rx+\Delta)]\nonumber\\
  =&E^{(R)}(x) + E^{(L)}(x), \label{eq:RL}
\end{align}
where $E^{(R)}(x)$ and $E^{(L)}(x)$ are the right-going and left-going 
waves, respectively.

We can build a standing wave component with $E^{(R)}(x)$ as
\begin{align}
  E^{(S)}(x)=&E^{(R)}(x) + [E^{(R)}(x)]^*\nonumber\\
  =& 2\Pi(x)\cos[K_rx+\Delta]\label{eq:RS}
\end{align}
and define the traveling wave component as the remaining part of the total field
\begin{align}
  E^{(T)}(x)=&E(x)-E^{(S)}(x)\nonumber\\
  =&E^{(L)}(x) - [E^{(R)}(x)]^*\nonumber\\
  =& [\Theta(x)-\Pi(x)]\exp[-i(K_rx+\Delta)].\label{eq:RT}
\end{align}
Hence, $2\Pi(x)$ and $[\Theta(x)-\Pi(x)]$ are the amplitudes of the standing 
wave and traveling wave components, respectively.
It is also possible to build a standing wave component with $E^{(L)}(x)$ as
\begin{align}
  E^{(S)}(x)=&E^{(L)}(x) + [E^{(L)}(x)]^*\nonumber\\
  =& 2\Theta(x)\cos[K_rx+\Delta]\label{eq:LS}
\end{align}
so that the traveling wave component reads
\begin{align}
  E^{(T)}(x)=&E(x)-E^{(S)}(x)\nonumber\\
  =&E^{(R)}(x) - [E^{(L)}(x)]^*\nonumber\\
  =& [\Pi(x)-\Theta(x)]\exp[i(K_rx+\Delta)].\label{eq:LT}
\end{align}

Comparing both ways of resolving the total field into its two components, we see
that in Eq. (\ref{eq:RT}) the traveling wave component is a left-going wave 
while in Eq. (\ref{eq:LT}) it is a right-going wave.
Hence, if in the expression of the field in Eq. (\ref{eq:RL}), the prevailing 
wave is the right-going wave $\Pi(x)\exp[i(K_rx+\varphi)]$ (i.e.,
$\Pi(x) > \Theta(x)$), we choose the standing and traveling wave components of
Eqs. (\ref{eq:LS}) and (\ref{eq:LT}).
In the opposite case of $\Pi(x) < \Theta(x)$, we choose the standing and 
traveling wave components of Eqs. (\ref{eq:RS}) and (\ref{eq:RT}).

Let us note that the imaginary part of the total field $E(x)$ is given in both
cases by
\begin{equation}
  \operatorname{Im}[E(x)] = [\Pi(x)-\Theta(x)]\sin[K_rx+\Delta].
\end{equation}
As expected, the presence of a traveling wave component, i.e., 
$|\Pi(x)-\Theta(x)|\ne 0$, makes $E(x)$ become complex instead of being real 
for a pure standing wave.



\end{document}